\documentclass[showpacs,aps,prd]{revtex4}
\input epsf

\begin{document}

\title{Could $Z_{b}(10610)$ be a $B^{*}\bar{B}$ molecular state?}
\author{Jian-Rong Zhang$^{1}$}
\author{Ming Zhong$^{1,2}$}
\author{Ming-Qiu Huang$^{1}$}
\affiliation{${}^{1}$ Department of Physics, National
University of Defense Technology, Hunan 410073, China\\
${}^{2}$ Kavli Institute for Theoretical Physics China,
CAS, Beijing 100190, China}

\begin{abstract}
Assuming the newly observed
structure $Z_{b}(10610)$ as a bottomonium-like molecular state $B^{*}\bar{B}$,
we calculate its mass in the framework of QCD sum rules.
The numerical result is $10.54\pm0.22~\mbox{GeV}$ for
$B^{*}\bar{B}$,
which coincide with the mass of $Z_{b}(10610)$. This
consolidates the statement made by Belle Collaboration that the
$Z_{b}(10610)$ resonance could be a $B^{*}\bar{B}$ molecular
state.
\end{abstract}
\maketitle

\section{Introduction}\label{sec1}

Very recently, Belle Collaboration observed two narrow structures
$Z_{b}(10610)$ and $Z_{b}(10650)$ in the
$\pi^{\pm}\Upsilon(nS)~~(n = 1, 2, 3)$ and $\pi^{\pm}h_{b}(mP)~~(m =
1, 2)$ mass spectra that are produced in association with a single
charged pion in $\Upsilon(5S)$ decays \cite{10620}. The measured
masses of the two structures are $10608.4\pm2.0~\mbox{MeV}$ and
$10653.2\pm1.5~\mbox{MeV}$, respectively. Experimental analysis favors quantum
numbers of $J^{P}=1^{+}$ for both states. As $Z_{b}(10610)$ and $Z_{b}(10650)$
are charged, they can not be simple $b\bar{b}$
combinations. The measured masses of these two new states exceed by only
a few $\mbox{MeV}$ the thresholds for the open beauty channels
$B^{*}\bar{B}$ and $B^{*}\bar{B}^{*}$. They
could be interpreted as molecular states and determined by the strong
interaction dynamics of $B^{*}\bar{B}$ and $B^{*}\bar{B}^{*}$ meson
pairs \cite{10620,10620-theory,10620-liu}.

The concepts of molecular states were put forward long ago in
\cite{long ago}.
Some of the exotic X, Y, and Z resonances have been
described as possible charmonium-like molecular candidates in the literatures
since their masses are very close to the meson-meson thresholds.
Explicitly, it is interpreted $Z^{+}(4430)$
as a $D^{*}\bar{D}_{1}$ molecular state
\cite{theory-Z4430}, $Y(3930)$ as a $D^{*}\bar{D}^{*}$ molecular
state \cite{theory-Y3930,Liu,3930-Ping}, $Y(4140)$ as a
$D_{s}^{*}\bar{D}_{s}^{*}$ molecular state \cite{Liu,theory-Y4140},
$Y(4260)$ as
a $\chi_{c}\rho^{0}$ \cite{Y4260-Liu} or
an $\omega\chi_{c1}$ molecular state \cite{Y4260-Yuan}, $X(4350)$ as a $D_{s}^{*}D_{s0}^{*}$ molecular state
\cite{X4350-Zhang,X4350-Ma},
$Y(4274)$ as a $D_{s}D_{s0}(2317)$ molecular state \cite{Y4274-Liu} etc..
If molecular states can be confirmed, QCD will be further
testified and then one will understand QCD low-energy behaviors more
deeply.

The newly observed $Z_{b}$ resonances may open a new window to study
molecular states in the bottomonium-like family. Therefore, it is
interesting to investigate whether they could be bottomonium-like
molecular candidates. The quantitative description of their
properties like masses are helpful for understanding their
structures. Unfortunately, quarks are confined inside hadrons in the
real world, and the strong interaction dynamics of these states are
governed by nonperturbative QCD effect completely. In this work, by
assuming $Z_{b}(10610)$ as a $B^{*}\bar{B}$ molecular state, we
calculate the mass of this resonance in the framework of QCD sum
rule (QCDSR) method \cite{svzsum}, which is a nonperturbative
formulation firmly rooted in QCD basic theory and has been used to
study some charmonium-like molecular states
\cite{SR-study,zgwang,zhang}. It is not so straightforward from the
meson-meson configuration of fields to construct a
$B^{*}\bar{B}^{*}$ current with a quantum number of $J^{P}=1^{+}$.
The $Z_{b}(10650)$ will not be discussed here. Our final numerical
result $10.54\pm0.22~\mbox{GeV}$ for $B^{*}\bar{B}$ agrees well with
the experimental data of $Z_{b}(10610)$, while the masses of the
$J^{P}= 1^{+}$ $b\bar{b}q\bar{q}$ tetraquark states were found to be
around $10.1\sim10.3~\mbox{GeV}$ in QCDSR
\cite{tetraquark-1,tetraquark-2}, which are lower than the measured
values of $Z_{b}$ states. The present work thus favors that the
$Z_{b}(10610)$ resonance could be a $B^{*}\bar{B}$ molecular state
rather than a $b\bar{b}q\bar{q}$ tetraquark state.

The rest of the paper is organized as three parts. We discuss QCD
sum rules for the molecular state in Sec. \ref{sec2}, where the
phenomenological representation and the operator product expansion (OPE) contribution up to
dimension six operators for the two-point correlator are derived.
The numerical analysis is made in Sec. \ref{sec3}. The mass of the
$B^{*}\bar{B}$ molecular state is extracted
out and found to coincide with the experimental value of
$Z_{b}(10610)$ resonance. The Sec. \ref{sec4} is
a short summary and outlook.

\section{molecular state QCD sum rules}\label{sec2}
The starting point of the QCD sum rule method is to construct the
interpolating current properly and then write down the correlator
(for reviews see \cite{overview1,overview2,overview3,overview4} and
references therein). The molecular state currents are built up with
the color-singlet currents for their composed hadrons.
As for $B^{*}\bar{B}$, the current is constructed as
\begin{eqnarray}
j^{\mu}_{B^{*}\bar{B}}&=&(\bar{q}_{e}\gamma^{\mu}b_{e})(\bar{b}_{f}i\gamma_{5}q_{f}),
\end{eqnarray}
where $q$ indicates the light quark and the subscript $e$ and $f$
are color indices. Note that the current is local and the four field
operators act at the same space-time point. It is a limitation
inherent in the QCDSR disposal of the hadrons since the bound states
are not point particles in a rigorous manner. The current is
different from that of tetraquark state which is diquark-antidiquark
configuration of fields. These two types of currents can be related
to each other by Fiertz rearrangement and differ by color and Dirac
factors \cite{overview-MN}. It will have a maximum overlap for the
molecular state with the meson-meson type current. The masses of the
$b\bar{b}q\bar{q}$ tetraquark states were calculated in QCDSR and
found to be around $10.1\sim10.3~\mbox{GeV}$ in \cite{tetraquark-1},
and $10144\pm106~\mbox{MeV}$ in \cite{tetraquark-2}. Both of the
predicted values for tetraquark states are lower than the $Z_b$ mass.
Therefore, it's hard to accommodate $Z_{b}$ as tetraquark states.

The two-point correlator is defined as
\begin{eqnarray}
\Pi^{\mu\nu}(q^{2})=i\int
d^{4}x\mbox{e}^{iq.x}\langle0|T[j^{\mu}_{B^{*}\bar{B}}(x)j^{\nu+}_{B^{*}\bar{B}}(0)]|0\rangle.
\end{eqnarray}
Lorentz covariance implies that it can be generally parameterized as
\begin{eqnarray}
\Pi^{\mu\nu}(q^{2})=(\frac{q^{\mu}q^{\nu}}{q^{2}}-g^{\mu\nu})\Pi^{(1)}(q^{2})+\frac{q^{\mu}q^{\nu}}{q^{2}}\Pi^{(0)}(q^{2}).
\end{eqnarray}
The term proportional to $g_{\mu\nu}$ will be chosen to extract the
mass sum rule. In phenomenology,
$\Pi^{(1)}(q^{2})$ can be expressed as
\begin{eqnarray}\label{ph}
\Pi^{(1)}(q^{2})=\frac{[\lambda^{(1)}]^{2}}{M_{B^{*}\bar{B}}^{2}-q^{2}}+\frac{1}{\pi}\int_{s_{0}}
^{\infty}ds\frac{\mbox{Im}\Pi^{(1)\mbox{phen}}(s)}{s-q^{2}},
\end{eqnarray}
where $M_{B^{*}\bar{B}}$ denotes the mass of the $B^{*}\bar{B}$ resonance,
$s_0$ is the threshold parameter, and $\lambda^{(1)}$ gives
the coupling of the current to the hadron $\langle0|j^{\mu}_{B^{*}\bar{B}}|B^{*}\bar{B}\rangle=\lambda^{(1)}\epsilon^{\mu}$.
In the OPE side, $\Pi^{(1)}(q^{2})$ can be written as
\begin{eqnarray}\label{ope}
\Pi^{(1)}(q^{2})=\int_{4m_{b}^{2}}^{\infty}ds\frac{\rho^{\mbox{OPE}}(s)}{s-q^{2}},
\end{eqnarray}
where the spectral density is
$\rho^{\mbox{OPE}}(s)=\frac{1}{\pi}\mbox{Im}\Pi^{\mbox{(1)}}(s)$.
Applying quark-hadron duality and making a Borel transform,
we have
the sum rule from Eqs. (\ref{ph}) and (\ref{ope})
\begin{eqnarray}\label{sr2}
[\lambda^{(1)}]^{2}e^{-M_{B^{*}\bar{B}}^{2}/M^{2}}&=&\int_{4m_{b}^{2}}^{s_{0}}ds\rho^{\mbox{OPE}}(s)e^{-s/M^{2}},
\end{eqnarray}
with $M^2$ the Borel parameter.
Making
the derivative in terms of $M^2$ to the sum rule and then dividing
by itself, we have the mass of the $B^{*}\bar{B}$ state
\begin{eqnarray}\label{sum rule 2}
M_{B^{*}\bar{B}}^{2}&=&\int_{4m_{b}^{2}}^{s_{0}}ds\rho^{\mbox{OPE}}s
e^{-s/M^{2}}/
\int_{4m_{b}^{2}}^{s_{0}}ds\rho^{\mbox{OPE}}e^{-s/M^{2}}.
\end{eqnarray}

For the OPE calculations, we work at leading order in $\alpha_{s}$
and considers condensates up to dimension six, with the similar
techniques developed in \cite{technique}. To keep the heavy-quark
mass finite, the momentum-space expression for the heavy-quark
propagator and the expressions with two and three gluons attached
are used \cite{reinders}. The light-quark part of the correlation
function is calculated in the coordinate space and then
Fourier-transformed to the momentum space in $D$ dimension. The
resulting light-quark part is combined with the heavy-quark part
before it is dimensionally regularized at $D=4$. The spectral
density can be written as
\begin{eqnarray}
\rho^{\mbox{OPE}}(s)=\rho^{\mbox{pert}}(s)+\rho^{\langle\bar{q}q\rangle}(s)+\rho^{\langle\bar{q}q\rangle^{2}}(s)+\rho^{\langle
g\bar{q}\sigma\cdot G q\rangle}(s)+\rho^{\langle
g^{2}G^{2}\rangle}(s)+\rho^{\langle g^{3}G^{3}\rangle}(s),
\end{eqnarray}
where $\rho^{\mbox{pert}}$, $\rho^{\langle\bar{q}q\rangle}$,
$\rho^{\langle\bar{q}q\rangle^{2}}$, $\rho^{\langle
g\bar{q}\sigma\cdot G q\rangle}$, $\rho^{\langle
g^{2}G^{2}\rangle}$, and $\rho^{\langle g^{3}G^{3}\rangle}$ are the
perturbative, quark condensate, four-quark condensate, mixed
condensate, two-gluon condensate, and three-gluon condensate
spectral densities, respectively. They are
\begin{eqnarray}
\rho^{\mbox{pert}}(s)&=&\frac{3}{2^{12}\pi^{6}}\int_{\alpha_{min}}^{\alpha_{max}}\frac{d\alpha}{\alpha^{3}}\int_{\beta_{min}}^{1-\alpha}\frac{d\beta}{\beta^{3}}(1-\alpha-\beta)(1+\alpha+\beta)r(m_{b},s)^{4},\nonumber\\
\rho^{\langle\bar{q}q\rangle}(s)&=&-\frac{3\langle\bar{q}q\rangle}{2^{7}\pi^{4}}m_{b}\int_{\alpha_{min}}^{\alpha_{max}}\frac{d\alpha}{\alpha^{2}}\int_{\beta_{min}}^{1-\alpha}\frac{d\beta}{\beta}(1+\alpha+\beta)r(m_{b},s)^{2},\nonumber\\
\rho^{\langle\bar{q}q\rangle^{2}}(s)&=&\kappa\frac{\langle\bar{q}q\rangle^{2}}{2^{4}\pi^{2}}m_{b}^{2}\sqrt{1-4m_{b}^{2}/s},\nonumber\\
\rho^{\langle g\bar{q}\sigma\cdot G q\rangle}(s)&=&\frac{3\langle
g\bar{q}\sigma\cdot G
q\rangle}{2^{8}\pi^{4}}m_{b}\int_{\alpha_{min}}^{\alpha_{max}}d\alpha\bigg\{\int_{\beta_{min}}^{1-\alpha}\frac{d\beta}{\beta}r(m_{b},s)-\frac{2}{1-\alpha}[m_{b}^{2}-\alpha(1-\alpha)
s]\bigg\},\nonumber\\
\rho^{\langle g^{2}G^{2}\rangle}(s)&=&\frac{\langle
g^{2}G^{2}\rangle}{2^{11}\pi^{6}}m_{b}^{2}\int_{\alpha_{min}}^{\alpha_{max}}\frac{d\alpha}{\alpha^{3}}\int_{\beta_{min}}^{1-\alpha}d\beta(1-\alpha-\beta)(1+\alpha+\beta)r(m_{b},s),\nonumber\\
\rho^{\langle g^{3}G^{3}\rangle}(s)&=&\frac{\langle
g^{3}G^{3}\rangle}{2^{13}\pi^{6}}\int_{\alpha_{min}}^{\alpha_{max}}\frac{d\alpha}{\alpha^{3}}\int_{\beta_{min}}^{1-\alpha}d\beta(1-\alpha-\beta)(1+\alpha+\beta)[r(m_{b},s)+2
m_{b}^{2}\beta],\nonumber
\end{eqnarray}
with $r(m_{b},s) = (\alpha+\beta)m_{b}^2 - \alpha \beta s$. The
integration limits are given by
$\alpha_{min}=\Big(1-\sqrt{1-4m_{b}^{2}/s}\Big)/2$,
$\alpha_{max}=\Big(1+\sqrt{1-4m_{b}^{2}/s}\Big)/2$ and
$\beta_{min}=\alpha m_{b}^{2}/(s\alpha-m_{b}^{2})$. We have applied
$\langle q\bar{q}q\bar{q}\rangle=\kappa\langle q\bar{q}\rangle^2$ to
estimate the four-quark condensate, where the parameter $\kappa$ is
introduced to account for deviation from the factorization
hypothesis \cite{overview2}.

\section{Numerical analysis}\label{sec3}
In this section, the Eq.(\ref{sum rule 2}) will be numerically
analyzed. The $b$ quark mass is taken as
$m_{b}=4.20^{+0.17}_{-0.07}~\mbox{GeV}$ \cite{PDG}. The condensates
are $\langle\bar{q}q\rangle=-(0.23\pm0.03)^{3}~\mbox{GeV}^{3}$,
$\langle g\bar{q}\sigma\cdot G
q\rangle=m_{0}^{2}~\langle\bar{q}q\rangle$,
$m_{0}^{2}=0.8~\mbox{GeV}^{2}$, $\langle
g^{2}G^{2}\rangle=0.88~\mbox{GeV}^{4}$, and $\langle
g^{3}G^{3}\rangle=0.045~\mbox{GeV}^{6}$ \cite{overview2,technique}.
In the QCDSR approach, there is approximation in the OPE of the
correlation function, and there is a very complicated and largely
unknown structure of the hadronic dispersion integral in the
phenomenological side. Therefore, the match of the two sides is not
independent of $M^{2}$.  One expects that there exists a range of
$M^{2}$, in which the two sides have a good overlap and the sum rule
can work well. In practice, one can analyse the convergence in the
OPE side and the pole contribution dominance in the phenomenological
side to determine the allowed Borel window: on one hand, the lower
constraint for $M^{2}$ is obtained by the consideration that the
perturbative contributions should be larger than the condensate
contributions, so that the convergence of the OPE is under control
and the higher dimension terms can be safely ignored; on the other
hand, the upper limit for $M^{2}$ is obtained by the restriction
that the pole contributions should be larger than the continuum
state contributions, so as to guarantee that the contributions from
high resonance states and continuum states remains a small part in
the phenomenological side. Meanwhile, the threshold parameter
$\sqrt{s_{0}}$ is not completely arbitrary but characterizes the
beginning of the continuum states. The energy gap between the
groundstate and the first excitation is around $500~\mbox{MeV}$ in
many cases of nucleons or charmonium-like states. Whereas,
this may not be always straightforwardly generalized to the bottomonium-like states
since their absolute masses are much larger than masses of nucleons or
charmonium-like states. One should consider whether there is a
proper region of Borel parameter $M^{2}$ for a fixing
$\sqrt{s_{0}}$, or whether one could find a working window for the
sum rule. On all accounts, it is
expected that the two sides have a good overlap in the determined
work window and information on the resonance can be safely
extracted.

Firstly, we take the factorization hypothesis of the four-quark
condensate $\langle q\bar{q}q\bar{q}\rangle=\kappa\langle
q\bar{q}\rangle^2$ with $\kappa=1$. The comparison between pole and
continuum contributions of sum rule (\ref{sr2}) as a function of the
Borel parameter $M^2$ for the threshold value
$\sqrt{s_{0}}=11.4~\mbox{GeV}$ is shown in FIG. 1. Its OPE
convergence by comparing the perturbative, quark condensate,
four-quark condensate, mixed condensate, two-gluon condensate, and
three-gluon condensate contributions as a function of $M^2$ is shown
in FIG. 2. The ratio of perturbative contributions to the total OPE
contributions at $M^{2}=9.0~\mbox{GeV}^{2}$ is nearly $58\%$ and it
increases with $M^{2}$. Thus the perturbative contributions will
dominate in the total OPE contributions when
$M^{2}\geq9.0~\mbox{GeV}^{2}$. On the other hand, the relative pole
contribution is approximate to $50\%$ at $M^{2}=10.5~\mbox{GeV}^{2}$
and it decreases with $M^{2}$. In order to guarantee that the pole
contribution can dominate in the total contributions, we have the
value $M^{2}\leq10.5~\mbox{GeV}^{2}$. Thus, the range of $M^{2}$ for
$B^{*}\bar{B}$ is taken as $M^{2}=9.0\sim10.5~\mbox{GeV}^{2}$ for
$\sqrt{s_0}=11.4~\mbox{GeV}$. Similarly, the proper range of $M^{2}$
is $9.0\sim9.8~\mbox{GeV}^{2}$ for $\sqrt{s_0}=11.2~\mbox{GeV}$, and
$9.0\sim11.2~\mbox{GeV}^{2}$ for $\sqrt{s_0}=11.6~\mbox{GeV}$. We
see that the corresponding Borel
parameter range is $M^{2}=9.0\sim9.1~\mbox{GeV}^{2}$ for $\sqrt{s_{0}}=11.0~\mbox{GeV}$, and
$M^{2}=9.0\sim9.4~\mbox{GeV}^{2}$ for
$\sqrt{s_{0}}=11.1~\mbox{GeV}$, which are very narrow as working
windows. Therefore, the threshold parameter $\sqrt{s_{0}}$ is taken
as $11.2\sim11.6~\mbox{GeV}$. The mass of
$B^{*}\bar{B}$ is numerically calculated to be $10.56\pm0.18~\mbox{GeV}$
and shown in FIG. 3.

The coupling constant $\lambda^{(1)}$ between current and the
particle is calculated from Eq.(\ref{sr2}) in the same working
windows. We arrive at $\lambda^{(1)}=0.27\pm0.07~\mbox{GeV}^{5}$.
The equivalent quantity of the tetraquark current is
$0.09\sim0.11~\mbox{GeV}^{5}$ in \cite{tetraquark-2}. The coupling
constant $\lambda^{(1)}$ is roughly three times as large as the
one of the tetraquark current and the meson-meson molecular current
has a larger overlap with the $Z_{b}$ state in comparison to the
diquark-antidiquark tetraquark current.

To investigate the effect of the factorization breaking, we assume
that $\langle q\bar{q}q\bar{q}\rangle=\kappa\langle
q\bar{q}\rangle^2$ with $\kappa=2$. From the similar analysis
process, the corresponding working windows are taken as:
$M^{2}=9.0\sim9.9~\mbox{GeV}^{2}$ for $\sqrt{s_0}=11.2~\mbox{GeV}$,
$M^{2}=9.0\sim10.6~\mbox{GeV}^{2}$ for $\sqrt{s_0}=11.4~\mbox{GeV}$,
and $M^{2}=9.0\sim11.4~\mbox{GeV}^{2}$ for
$\sqrt{s_0}=11.6~\mbox{GeV}$. We extract the mass value
$10.52\pm0.21~\mbox{GeV}$. Finally, we average two results for
$\kappa=1,2$ and arrive at the mass value $10.54\pm0.22~\mbox{GeV}$
for $B^{*}\bar{B}$, which agrees with the experimental value
$10608.4\pm2.0~\mbox{MeV}$ for $Z_{b}(10610)$.

\begin{figure}
\centerline{\epsfysize=6.0truecm\epsfbox{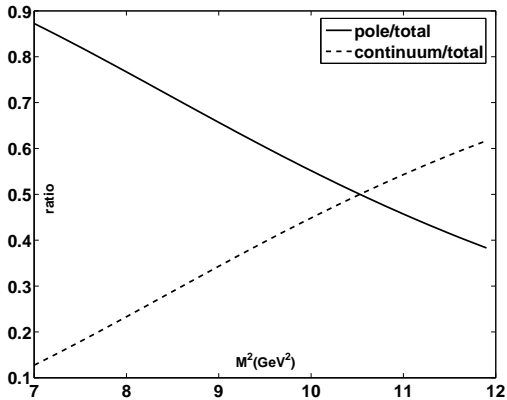}}
\caption{The phenomenological contribution in sum rule
(\ref{sr2}) for $\sqrt{s_{0}}=11.4~\mbox{GeV}$.
The solid line is the relative pole contribution (the pole
contribution divided by the total, pole plus continuum contribution)
as a function of $M^2$ and the dashed line is the relative continuum
contribution.}
\end{figure}

\begin{figure}
\centerline{\epsfysize=6.0truecm\epsfbox{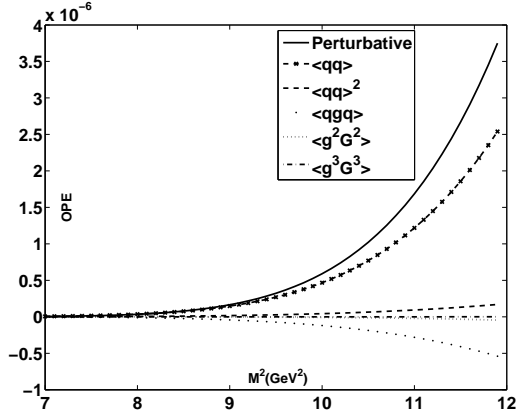}}
\caption{The OPE contribution in sum rule
(\ref{sr2}) for $\sqrt{s_{0}}=11.4~\mbox{GeV}$.
The OPE convergence is shown by comparing the
perturbative, quark condensate, four-quark condensate, mixed condensate, two-gluon condensate, and three-gluon
condensate contributions.}
\end{figure}

\begin{figure}
\centerline{\epsfysize=6.0truecm
\epsfbox{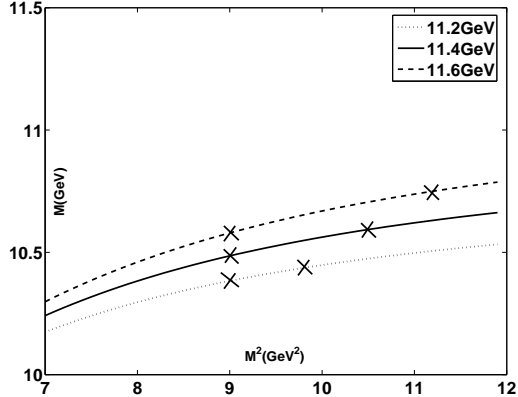}}\caption{
The mass of the $B^{*}\bar{B}$ molecular state as
a function of $M^2$ from sum rule (\ref{sum rule 2}). The continuum
thresholds are taken as $\sqrt{s_0}=11.2\sim11.6~\mbox{GeV}$. The
ranges of $M^{2}$ is $9.0\sim9.8~\mbox{GeV}^{2}$ for
$\sqrt{s_0}=11.2~\mbox{GeV}$, $9.0\sim10.5~\mbox{GeV}^{2}$ for
$\sqrt{s_0}=11.4~\mbox{GeV}$, and $9.0\sim11.2~\mbox{GeV}^{2}$ for
$\sqrt{s_0}=11.6~\mbox{GeV}$.}
\end{figure}

\section{Summary and outlook}\label{sec4}
By assuming $Z_{b}(10610)$ as a $B^{*}\bar{B}$ molecular state, the QCD sum rule
method has been applied to calculate the mass of the resonance.
Our numerical result is $10.54\pm0.22~\mbox{GeV}$ for
$B^{*}\bar{B}$.
It is compatible with the newly measured experimental data of
$Z_{b}(10610)$ by Belle Collaboration, which
supports the statement that $Z_{b}(10610)$
resonance could be a $B^{*}\bar{B}$ molecular state. It is expected that this work is helpful
for understanding the structure of $Z_{b}(10610)$.
For the newly observed structure $Z_{b}(10650)$, one could consider
how to construct a $B^{*}\bar{B}^{*}$ molecular state current with a quantum number of
$J^{P}=1^{+}$ from the meson-meson configuration
of fields.
For further work,
one needs to take into
account other dynamical analysis to identify the structures of $Z_{b}$ hadrons.

\begin{acknowledgments}
This work was supported in part by the National Natural Science
Foundation of China under Contract Nos.11105223, 10947016 and 10975184.
\end{acknowledgments}

\end{document}